\documentclass[journal,twoside,web]{ieeecolor}
\usepackage{jsen1}
\usepackage{cite}
\usepackage{amsmath,amssymb,amsfonts}
\usepackage{algorithmic}
\usepackage{graphicx}
\usepackage{textcomp}
\usepackage{wrapfig}

\usepackage[utf8]{inputenc}
\usepackage[T1]{fontenc}
\usepackage{mathptmx}
\usepackage{etoolbox}
\usepackage{amssymb,amsmath,bm,textcomp,gensymb,color,soul}
\usepackage{graphicx,float,subfig}
\usepackage{lineno}

\definecolor{abstractbg}{rgb}{1,1,1}
\begin{document}
\title{Capacitive temperature sensing via displacement amplification}
\author{
Semih Taniker, Vincenzo Costanza, Paolo Celli, and Chiara Daraio

\thanks{S. Taniker was with the Division of Engineering and Applied Science, California Institute of Technology, Pasadena, CA 91125, USA. }
\thanks{V. Costanza is with the Division of Engineering and Applied Science, California Institute of Technology, Pasadena, CA 91125, USA. }
\thanks{P. Celli was with the Division of Engineering and Applied Science, California Institute of Technology, Pasadena, CA 91125, USA. He is 
now with the Department of Civil Engineering, Stony Brook University, Stony Brook, NY 11794, USA.}
\thanks{C. Daraio is with the Division of Engineering and Applied Science, California Institute of Technology, Pasadena, CA 91125, USA (e-mail: daraio@caltech.edu). }
\thanks{The data that support the findings of this study are available from the corresponding author upon reasonable request.}
}
\IEEEtitleabstractindextext{%
\fcolorbox{abstractbg}{abstractbg}{%
\begin{minipage}{\textwidth}%
\begin{wrapfigure}[12]{r}{3in}%
\includegraphics[width=3in]{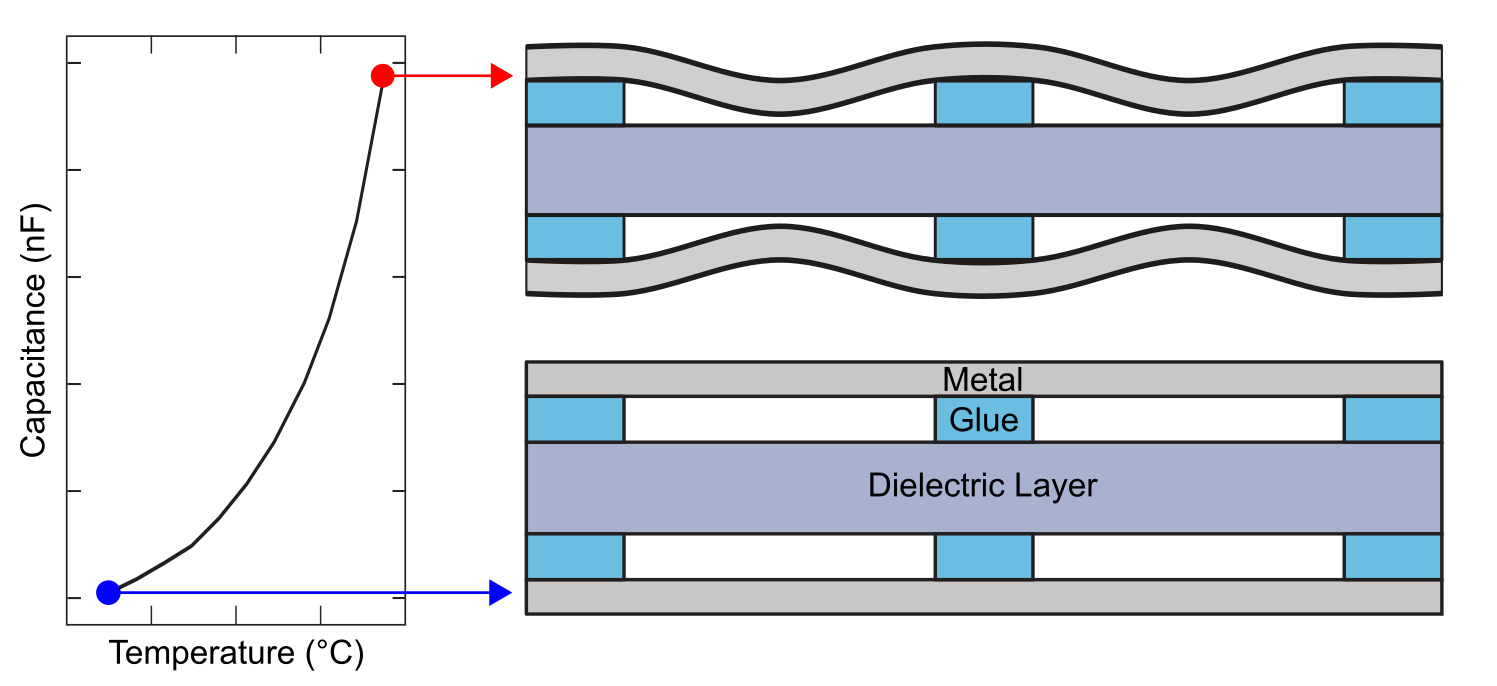}%
\end{wrapfigure}%
\begin{abstract}
We propose the realization of capacitive temperature sensors based on the concept of displacement amplification. Our design features two high coefficient of thermal expansion (CTE) metallic layers separated by a low-CTE dielectric layer; conductive and dielectric layers are then separated by a thin air gap and glued together at a few locations. As the temperature increases, the high-CTE layer tends to expand more than the low-CTE one. Owing to the constraint to planar expansion imposed by the low-CTE layer, this results in large out-of-plane displacements of the high-CTE layer -- hence the displacement amplification term. In our case, the high-CTE layer buckles and causes a reduction of the gap between conductive and dielectric layers; in turn, this results in a large change of capacitance. First, we illustrate the concept via numerical simulations. Then, we realize a low-cost prototype of such sensor by using aluminum foil as conductor, paper as dielectric, and by gluing the layers together with cyanoacrylate. Our results demonstrate the potential of this simple design as a route towards efficient and low-cost temperature sensors.
\end{abstract}

\end{minipage}}}

\maketitle

\section{Introduction}

Temperature sensors play a fundamental role in countless technologies and industrial applications. The most common type of temperature sensors measure heat transported to the detector through conductive/convective transport. Examples of such sensors are thermocouples, resistance temperature detectors, thermistors and Si- and Ge- semiconductors embedded in integrated circuits~\cite{feteira2009negative,becker1946properties}. Most of these technologies rely on changes in resistance of some components/materials due to changes in temperature. Few instances of capacitive sensors also exist and display sensitivities that are about 1\%/$\degree \mathrm{C}$ lower than those of more established sensing technologies~\cite{oertel1994capacitive}. Unlike contact temperature sensors, IR thermal sensors measure the temperature transported to the detector through infrared radiation. They are attractive because they are non-contact and have the potential to acquire information and detect events even in low-light and poor visual conditions. Among the different IR detectors, bolometers are extremely interesting for their commercial applications. In a bolometer, the incident IR radiation is absorbed by a black body, which heats up due to the absorbed energy. This change in temperature is then measured by a thermally-responsive layer as a change in resistivity or dielectric constant. Significant efforts have been dedicated to develop and optimize temperature resistive layers for microbolometers, as they represent the principal bottleneck for their sensitivity. Nowadays, the most employed microbolometers are VOx ~\cite{han2005vox,lv2007vox, kumar2009vox}, Si~\cite{syllaios1999amorphous} and Si-Ge alloys ~\cite{unold1993sigealloys,garcia2004sigealloys}, that display a temperature coefficient of resistance (TCR) between ~ 2\%/$\degree \mathrm{C}$ and 5\%/$\degree \mathrm{C}$.

Temperature sensors that rely on electromechanical principles -- where the electrical response stems from engineered arrangements of materials that yield desired deformations and changes in resistance/capacitance/dielectric constant -- are not common. Examples are capacitive sensors consisting of multi-layer cantilevers that bend in response to differential thermal expansion in each layer~\cite{ma2009micromachined, Ma2010, peroulis2015highly}. The bending deformations cause changes in the dielectric constant of one of the inner layers, leading to capacitance variations that can be detected by a readout circuit. However, these technologies are hampered by the fact that the strains achievable in a cantilever are limited, and the changes in dielectric constants with strains are small. Nonetheless, they find applications in MEMS, for example, due to their low energy consumption and low heat dissipation.

In recent years, considerable attention has been devoted to engineering structures with tailored response to environmental stimuli, such as temperature~\cite{sigmund1996,lakes2007,jefferson_tailorable_2009,wei2016jmps,wei2018,li_hoberman-sphere-inspired_2018}. Through geometric arrangements of materials with contrasting coefficient of thermal expansion (CTE), it is possible to obtain composite structures with extreme values of effective CTE. This includes, for example, structures with zero or negative effective CTE~\cite{gdoutos2013, yamamoto2014, wang2016}, or structures that can achieve extreme shape changes in response to temperature variations~\cite{boatti2017, tang2017, xu2018, liu2019encoding}. In recent work, some of the authors of this letter demonstrated that joining a high-CTE frame with a low-CTE bar allows to produce large tangential displacements of the outer frame~\cite{taniker2020temperature}, owing to displacement-amplification mechanisms~\cite{idogaki_bending_1996}.

In this letter, we leverage similar displacement amplification mechanisms to create capacitive temperature sensors. We start from thin layers of materials, one conductive and one dielectric, with drastically-different thermal expansion. We show that, by engineering their spatial arrangement as to cause large relative displacements between conductive and dielectric layers, we produce large changes in capacitance per degree, of about 600\,\% over $65\degree \mathrm{C}$. Our demonstration is carried out with low-cost sensors fabricated by manually bonding paper and aluminum foil with cyanoacrylate glue. However, especially considering that similar displacement-amplifying mechanisms are common in MEMS~\cite{Iqbal2019}, our sensors are scalable and could find applications in consumer electronics.  

\section{Unit cell design}

\begin{figure}[h]
\centering
\includegraphics[scale=0.47]{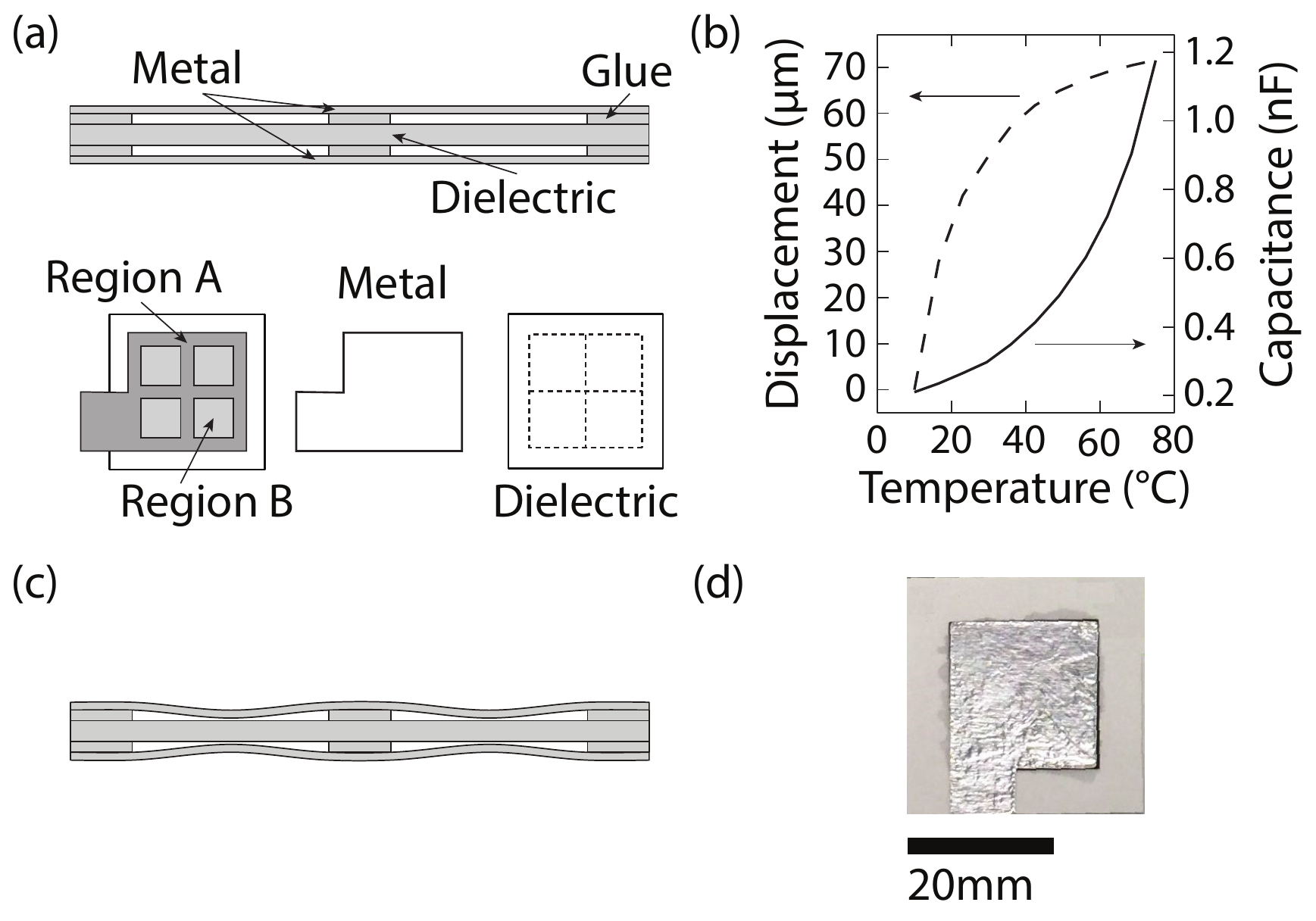}
\caption{(a) Schematic representation of the temperature sensing unit, with details of the various layers indicating regions A (glue, dielectric and metallic layers) and B (air, dielectric and metallic layers). (b) Average vertical displacement (dashed curve) and capacitance (continuous curve) vs.\ temperature of a unit featuring the dimensions written in the main text. (c) Schematic representation of the temperature sensing unit after a temperature increase. (d) A top view of the actual three-dimensional temperature sensing unit.}
\label{f:NumericalUnit}
\end{figure}

A sketch of a cross-section of the displacement-amplifying structure used to create our sensors is shown in Fig.~\ref{f:NumericalUnit}(a). It features two outer metallic layers of high-CTE material (aluminum foil) and an inner low-CTE, dielectric layer (paper). The layers are bonded with cyanoacrylate glue at selected locations, that are chosen to create large air gaps for the metallic layers to expand. As we increase temperature, the high-CTE conductive layers tend to expand (axially) more than the dielectric one. However, this axial expansion is limited by the presence of the low-CTE dielectric and by the fact that the layers are partially glued together~\cite{taniker2020temperature}. Thus, the conductive layers buckle and bulge tangentially (Fig.~\ref{f:NumericalUnit}(c)). The tangential displacement, in particular, is much larger than any axial displacement in the layers -- hence the displacement amplification term. This amplified bulging can either be directed inwards or outwards. In our case, the fabrication process biases the outer layers to displace towards the inner one, and to close the air gaps. In turn, this causes an increase in capacitance.

\section{Theoretical analysis}
\label{s:Analy}

Our first step is to perform numerical simulations on the two-dimensional version of our sensor shown in Fig.~\ref{f:NumericalUnit}(a). To do so, we resort to the commercial finite element (FE) platform Abaqus/Standard. We select 8-node bi-quadratic plane strain quadrilateral elements to perform the analysis, and we consider geometric nonlinearities. The inner, low-CTE, dielectric layer is modeled as a beam of length $l_d=20\,\mathrm{mm}$ and thickness $t_d=50\,\mathrm{\mu m}$, and is made of paper, with Young's modulus $E_L=3\,\mathrm{GPa}$, Poisson's ratio $\nu_L=0.25$, coefficient of thermal expansion $\alpha_L=4\,\,\mathrm{10^{-6}/\degree C}$). The outer high-CTE layers are here represented by beams of length $l_c=20\,\mathrm{mm}$ and thickness $t_c=12\,\mathrm{\mu m}$, and are made of Aluminum foil, with $E_H=70\,\mathrm{GPa}$, $\nu_H=0.33$, and $\alpha_H=23.1\,\,\mathrm{10^{-6}/\degree C}$)~\cite{ashby2016}. Our model also includes layers of glue, represented as blocks of length $l_g=2\,\mathrm{mm}$; since we have little control over the glue thickness, we fit this parameter to match our experimental results. The most accurate fit yields $t_g=71.56\,\mathrm{\mu m}$. Finally, we assume the coefficients of thermal expansion to be constant over large temperature ranges. 

The simulation results for a single unit are given in Fig.~\ref{f:NumericalUnit}(b). We choose $T_i=10 \degree \mathrm{C}$ as the initial temperature, while the final one is $T_f=75 \degree \mathrm{C}$. As the temperature rises, the metallic layer expands tangentially; its displacement is shown by the dashed curve. We can appreciate that these displacements are of the order of tens of $\mu m$ and are large in comparison to the individual layer thicknesses.

Obtaining an accurate measure of the change of capacitance as a function of temperature in our simulations is no simple task, as it requires us to understand the relationship between capacitance and deformation. There are two different cross-sections in our capacitive temperature sensors: one, region A, consists of glue, dielectric and metal layers; the other one, region B, consists of dielectric, metal layers and the air gap. It is important to note that the paper absorbs some of the glue after it is applied, and this causes its dielectric constant to be higher than the nominal value for paper-only. The characterization of the dielectric constant of the glue-impregnated paper is provided in the supplementary material (SM).

In order to obtain the temperature-dependent capacitance of this sensor from our simulations, we need to calculate the capacitance of regions A and B (Fig. \ref{f:NumericalUnit}(a)), and add them up. The capacitance of Region A (made of glue, dielectric and metal layers) is assumed to be constant and is calculated as
\begin{equation}
C_{A}=\frac{C_{g}\,C_{dA}}{C_{g}+C_{dA}},
\label{e:Cconstant}
\end{equation}
where
\begin{equation}
C_{g}=\frac{\kappa_{g}\,\varepsilon_0\,A_{A}}{2t_{g}},\quad C_{dA}=\frac{\kappa_{d}\,\varepsilon_0\,A_{A}}{t_d}.
\label{e:Cglue}
\end{equation}
The capacitance of the Region B is temperature dependent and it is calculated as
\begin{equation}
C_{B}(T)=\frac{C_{air}(T)\,C_{dB}}{C_{air}(T)+C_{dB}},
\label{e:CB}
\end{equation}
where
\begin{equation}
C_{air}(T)=\frac{\kappa_{air}\,\varepsilon_0\,A_{B}}{2t_{air}(T)},\quad C_{dB}=\frac{\kappa_{d}\,\varepsilon_0\,A_{B}}{t_{d}}
\label{e:Cair}
\end{equation}
and
\begin{equation}
t_{air}(T)=t_{g}-\frac{1}{n}\displaystyle\sum_{i=1}^{n} d_i(T).
\label{e:hAir}
\end{equation}
In Eqs.~\ref{e:Cconstant}-\ref{e:hAir}, $C_A$ is the capacitance of Region A, $C_B$ is the capacitance of Region B, $A_{A}$ is the cross-sectional area of region A, $A_{B}$ is the area of region B, $\kappa_{g}$ is the relative permittivity of glue, $\kappa_{air}$ is the relative permittivity of air, $\kappa_{d}$ is the relative permittivity of the dielectric layer (glue-impregnated paper),  $\varepsilon_0$ is the vacuum permittivity, $t_{g}$ is the thickness of the glue layer, $t_{d}$ is the thickness of the dielectric layer, $n$ is the number of nodes corresponding to the lower surface of the metallic layer in the FEM model, and $d_i$ is the displacement of each node. Finally, the total temperature-dependent capacitance of the sensor can be calculated as $C_{total}(T)=C_{A}+C_{B}(T)$.

Using the numerical displacement results shown in Fig.~\ref{f:NumericalUnit}(b) and this definition of total capacitance, we plot the variation of capacitance as a function of temperature in Fig.~\ref{f:NumericalUnit}(b). We can see that the displacement of the conducting layers, as previously claimed, is accompanied by large changes in capacitance, of the order of 1000\,pF over a range of $65 \degree \mathrm{C}$. To validate our numerical predictions, we manufacture and test a set of paper-aluminum sensors, fabricated as described in Appendix \ref{a:Methods}.

\section{Experimental validation}
\label{s:Exp}

Capacitors are nowadays fabricated using different approaches (e.g., electrolytic, ceramic, tantalum, etc.). Manufacturers provide various figures of merit to compare the performances of different components. The most important are the frequency behavior, the operating voltage and the temperature dependence. The frequency dependence of a capacitor provides its effectiveness when used in decoupling supplies from the input noise. For this study, measuring the frequency behavior of our sensors is important for two reasons: (1) at the operating frequency, the sensors has to be away from their resonances to avoid unwanted oscillations; and (2) their parasitic resistance and inductance have to be negligible. Therefore, we measure the change of capacitance as a function of frequency in our sample, to select the most effective operating frequency for the readout system (Fig.~\ref{f:freq_cap}(a) of the Appendix). The drop in the capacitance for increasing frequencies is likely caused by the decrease of the paper dielectric constant, as relaxation processes arise in the paper matrix \cite{simula1999measurement}. In principle, since there is no resonance in the investigated frequency range, we could have chosen any frequency. However, in order to decrease the effect of parasitics on the measurement, we target the lower frequency spectrum (20 – 50 Hz) to electrically stimulate the sensor. Given the frequency dependence of the capacitance, we measure the dielectric constant through impedance spectroscopy (Fig.~\ref{f:freq_cap}(b) of the Appendix) in order to fit the experimental results and the numerics. The static permittivity was measured to be 20. However, given the effect of frequency on the capacitance we used the value of 15.81 to fit the experimental data with the numerical simulation. After this study, we finalized the design of the capacitance meter based on the measurement of the phase lag of a first order RC circuit (Fig.~\ref{f:meas_circuit} of the SM), which was then implemented on a printed circuit board.

The second figure of merit for the capacitive temperature sensor is the operating voltage. The magnitude of the applied field can break the dielectric, or lower the permittivity, by influencing the spontaneous polarization of the capacitor. When measuring the dependence of the capacitance to the applied voltage (100\,mVpp, 250\,mVpp and 500\,mVpp) at different frequencies, we found no differences.

\begin{figure}[h]
\centering
\includegraphics[scale=0.47]{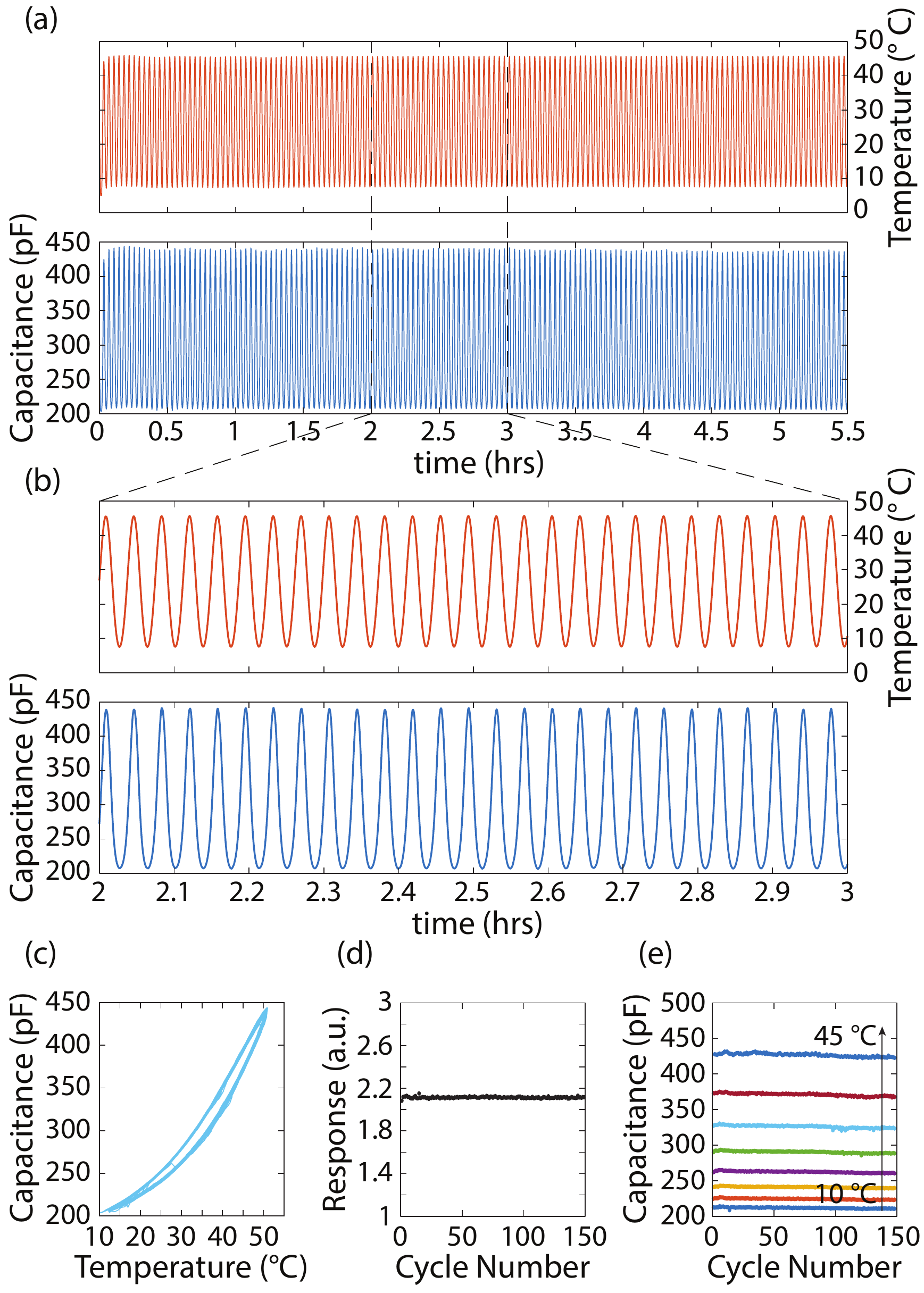}
\caption{Cyclic response of the capacitive temperature sensor. (a) Time evolution of the temperature and capacitance over 5.5 hours. (b) Details of the recorded signals, showing some nonlinearity in the capacitance curve. (c) Capacitance evolution as a function of temperature, indicating nonlinearity and histeretic behavior. (d) Evolution of the ratio between capacitance at the highest ($45\degree \mathrm{C}$) and lowest ($8\degree \mathrm{C}$) testing temperatures, indicating the stability of our measurements. (e) Capacitance vs.\ cycle number, for various temperatures, also indicating stability in our measurements.}
\label{f:time_response}
\end{figure}

Lastly, we analyze the change of the capacitance as a function of temperature. Standard capacitors are built to have the least change of capacitance with temperature change. In this work, since the paper capacitors are used as temperature sensors, our objective is to obtain a change in capacitance as large as possible for a given $\Delta T$. To assess the performance of the sensors, we cycle the temperature in their environment from $8\degree \mathrm{C}$ to $45\degree \mathrm{C}$ for over 5 hours (Fig.~\ref{f:time_response}(a)). 
The capacitance is acquired by the capacitance meter at $44\,\mathrm{S/s}$. Over this temperature range, the capacitance is measured to vary between $\sim$200\,pF to $\sim$450\,pF. As it can be seen in a zoomed-in section of the cycles (Fig.~\ref{f:time_response}(b)) and in the capacitance versus temperature plot (Fig.~\ref{f:time_response}(c)), the relation between temperature and capacitance is nonlinear. In this scenario, a linear temperature coefficient  of resistance (TCR) cannot be defined. Therefore, we define the sensor’s temperature response as the ratio between the capacitance at the highest temperature ($45\degree \mathrm{C}$) divided by the capacitance at the lowest temperature ($8\degree \mathrm{C}$) of the temperature cycle. The response is shown in figure Fig.~\ref{f:time_response}(d) as a function of the cycle number. The capacitance more than doubles over a $37\degree \mathrm{C}$ interval and is stable over time. The stability over time is also confirmed by Fig.~\ref{f:time_response}E, where the capacitance at different temperatures ($10\degree \mathrm{C}$ to $45\degree \mathrm{C}$ in $5\degree \mathrm{C}$ increments) is plotted versus the cycle number. The capacitance values measured at different temperatures are constant over time.

\begin{figure}[H]
\centering
\includegraphics[scale=0.49]{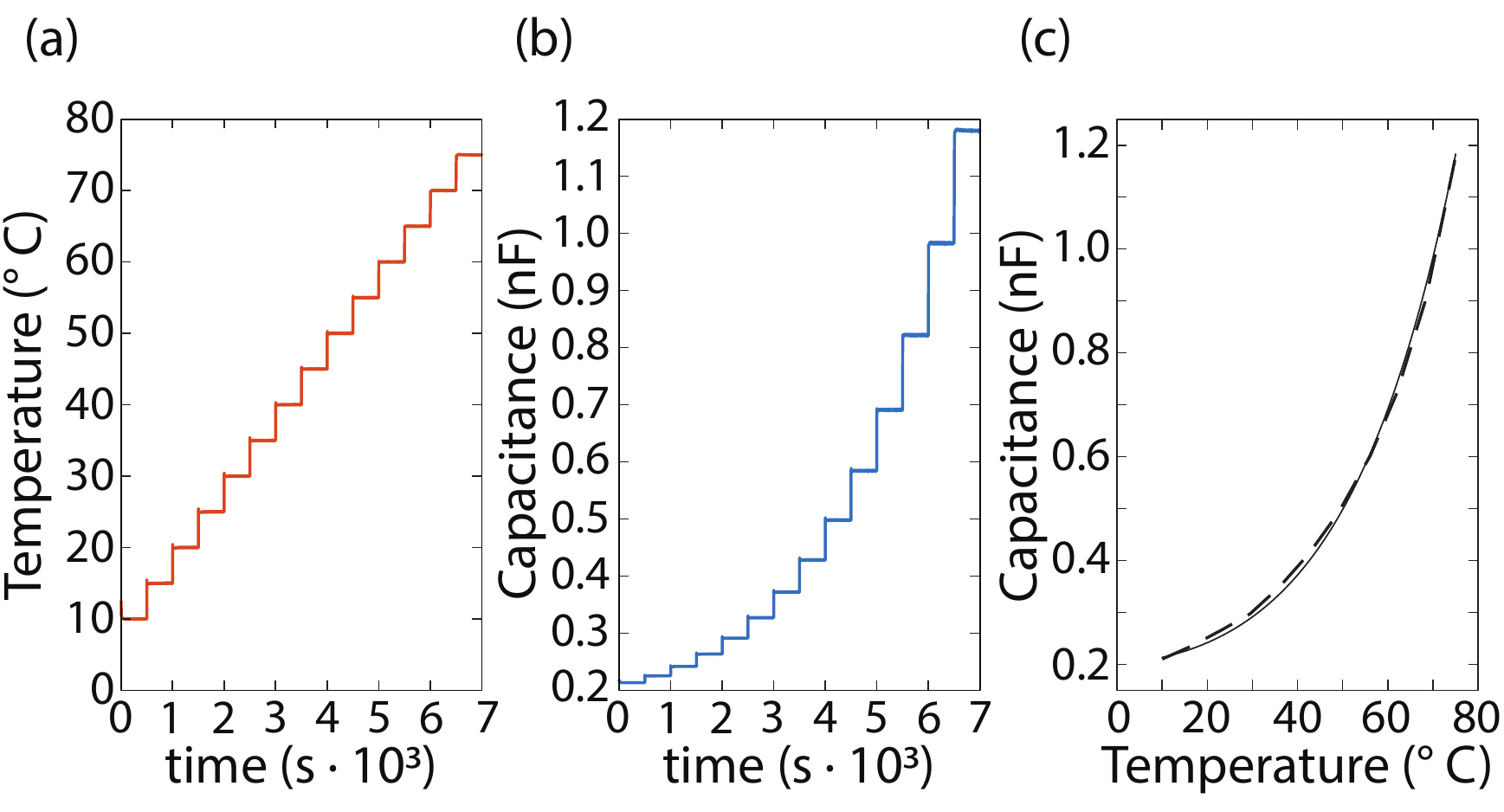}
\caption{Comparison of experimental  and numerical results. (a) Temperature evolution law used to produce results that are comparable to the numerical curve. (b) Capacitance read-out as a result of the temperature protocol shown in a). (c) Comparison between the numerical capacitance vs.\ temperature curve and the experimental one, obtained by fitting a line to the curve in (b).}
\label{f:time_temp_capacitance}
\end{figure}

As shown in Fig.~\ref{f:time_response}(c),  due to the thermal inertia of the system, the capacitance versus temperature shows hysteretical behavior, which makes it challenging to effectively compare the experimental and numerical results. Therefore, we characterize the system with temperature steps instead of temperature cycles. By doing so, we allow the temperature on the samples to settle, to account for the thermal inertia of the system and the consequent time, achieving a quasi-static regime. The temperature is controlled via a PID controller and raised from $10\degree \mathrm{C}$ to $75\degree \mathrm{C}$ in steps of $5\degree \mathrm{C}$, and held costant for 10 minutes (Fig.~\ref{f:time_temp_capacitance}(a)). 
As in Fig.~\ref{f:time_response}, the capacitance closely follows the temperature, going from $\sim$200\,pF at $10\degree \mathrm{C}$ to 1200\,pF at $75\degree \mathrm{C}$, and shows that the sample is stable over time (Fig.~\ref{f:time_temp_capacitance}(b)). The capacitance versus temperature curve measured following this procedure is then compared to the numerical results (Fig.~\ref{f:time_temp_capacitance}(c)), finding excellent agreement. 

To test the functionality of the capacitive sensor in a real-world scenario, we use the paper capacitors as IR absorbers. We fabricate a standard paper capacitor following the procedure described in Appendix \ref{a:Methods}, sandwiching a thin piece of paper between two aluminum layers. We apply black paint on top of one of the two aluminum layers to create a surface that absorbs the incident IR radiation. As the radiation is absorbed by the black paint, the temperature of the absorbing layer increases, transferring the heat from the absorbing layer to the aluminum foil; this, in turn, alters the capacitance of the sensor. The sensor is kept on a thermoelectric cooler and a PID controller, in order to keep its temperature constant and to isolate the effects of the absorbed radiation. The radiation source is turned ON and OFF periodically for 10 ms, and the capacitance is measured using the readout systems described in the Appendix \ref{a:cte}(Fig.~\ref{f:response_time}(a)).
\begin{figure}[H]
\centering
\includegraphics[scale=0.49]{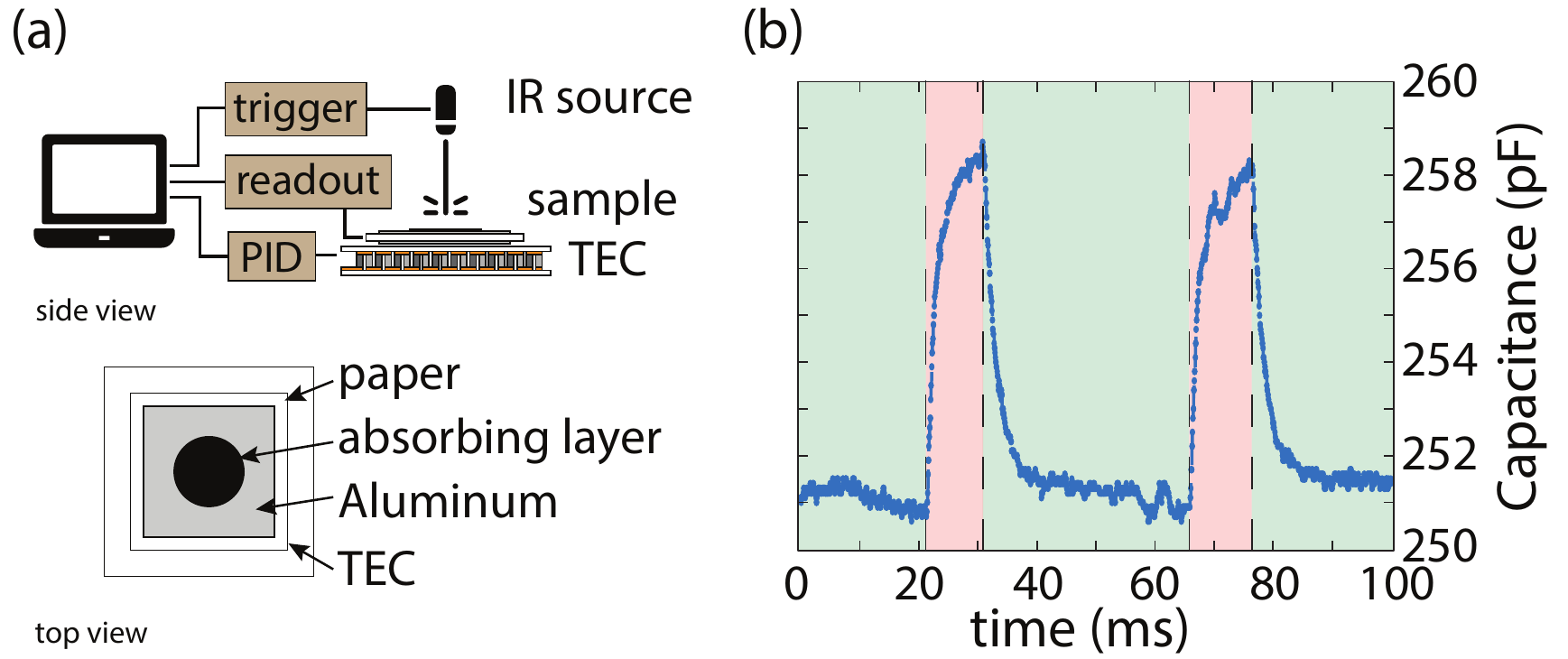}
\caption{Response of our capacitive temperature sensor to IR. (a) Experimental setup for testing. (b) Capacitance vs.\ time. The shaded red and green regions indicate when the IR source is ON or OFF, respectively.}
\label{f:response_time}
\end{figure}
From the time evolution of the measured capacitance (Fig.~\ref{f:response_time}(b)), we can see that this is constant when the IR source is off, while an increase in capacitance is registered when turning on the IR source. we can also see that the sensor recovers after each ON/OFF cycle. 

\section{Conclusion}

In this work, we have demonstrated that it is possible to create high-performing capacitive temperature sensors by leveraging mechanical displacement-amplification mechanisms. We have also shown that these sensors can be used for IR sensing applications. The proposed mechanism is suitable for micro-manufacturing and for the creation of low-cost IR sensors. For example, the fabrication of suspended Al membranes on top of a $Si$ base and a $SiO_2$ dielectric layer is a possible approach~\cite{guedes2011aluminum}. 

Micro-bolometers cover more than 95\% of the uncooled IR detectors market ~\cite{rogalski2012history}. The most common sensitive layer used in micro-bolometers is vanadium oxide, having a coefficient of resistance at room temperature of more than 4\% per degree~\cite{wang2013nanostructured,zia2016synthesis}. Compared to uncooled micro-bolometers, the proposed micro-mechanism can achieve considerably higher sensitivity. Moreover, spatial arrays of our sensors could be used to create temperature sensing surfaces with finite pixelation, that could find use as low-cost temperature mapping devices in applications where monitoring the temperature is crucial, such as to monitor the health of batteries for electric vehicles. 



\appendices

\section{Materials and Methods }
\label{a:Methods}

To validate our numerical and theoretical predictions, we manufacture and test a set of bi-material specimens. We use desktop cutting machine (Silhouette Cameo 3) on 50 um-thick sheets of copy paper and 12 um-thick aluminum foil, and we manually glue the Al foils to the paper. The samples were connected to the electronics read-out through copper wires attached via copper tape. 
The capacitance was measured with three independent systems an LCR meter (BK891, B\&K Precision) to measure the capacitance frequency dependence, a lock-in amplifier to measure the static permittivity (MFIA 5 MHz, Zurich Instruments) (see Section \ref{a:freq}), and a custom-made capacitance - meter system to measure the temperature – capacitance dependence (see Appendix B). The temperature cycling was generated via a peltier element controlled by a specifically designed control PID circuit and measured via a platinum sensor, previously calibrated on a thermal camera (FLIR A655sc).  

\section{Capacitance Characterization}
\label{a:freq}

To measure the electrical properties of the fabricated capacitors we measured the capacitance as a function of frequency. We swept the frequency from 20Hz to 10kHz, measuring 1000 different frequencies with the LCR Meter. For each frequency we acquired 50 points and calculated the average of these points. The static permittivity was calculated from the conductivity extracted though an impedance spectrum measured with the lock in amplifier. The real and imaginary part for the conductivity was measured between 100 mHz and 1 MHz at 1V. Then the real and imaginary part of the permittivity was respectively calculated by:

\begin{equation}
\varepsilon{'}=\frac{1}{2 \pi f \varepsilon_{0} \sigma''}
\label{e:epsilon1}
\end{equation}

\begin{equation}
\varepsilon{''}=\frac{1}{2 \pi f \varepsilon_{0} \sigma'}
\label{e:epsilon2}
\end{equation}

The static permittivity corresponds to the low frequency plateau of the real permittivity. 

\begin{figure}[!htb]
\centering
\includegraphics[scale=0.5]{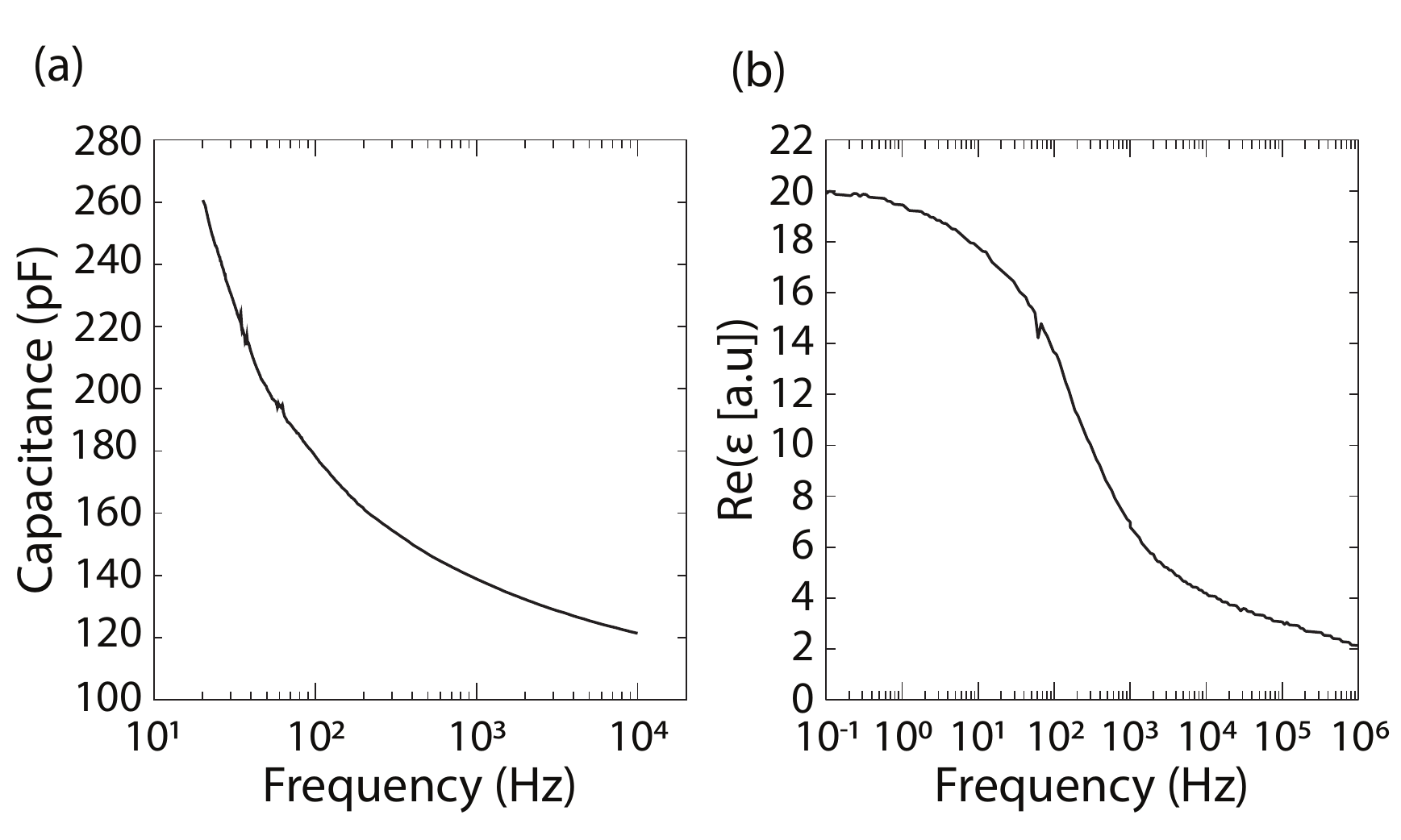}
\caption{Frequency dependence of the capacitive temperature sensors.}
\label{f:freq_cap}
\end{figure}

\section{Readout System}
\label{a:cte}

\begin{figure}[H]
\centering
\includegraphics[scale=0.49]{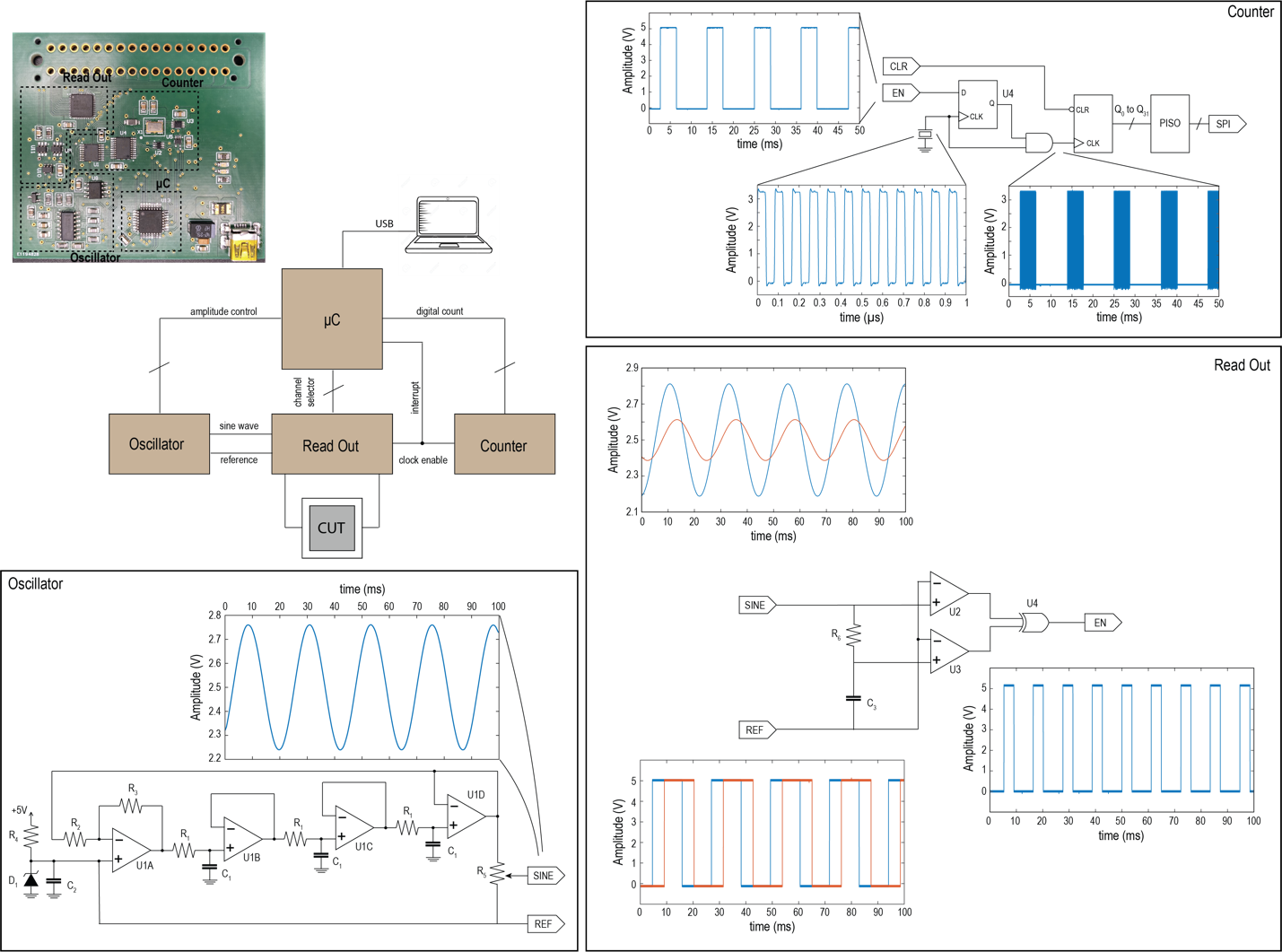}
\caption{Capacitive measurement circuit designed to measure temperature-capacitance relation precisely.}
\label{f:meas_circuit}
\end{figure}

The custom-made electronics to measure the capacitance of the sensor is based on the measurement of the phase lag in a RC circuit. Generally speaking, a first order low pass RC filter shows a drop in the response of -20dB/dec after the cut-off frequency and a phase lag that reaches -45° at the cut-off, while it is 0° one decade before and -90° one decade after. At a constant frequency and resistance, a change in the capacitance will result in a change in the phase lag between the input and the output signals, according to the relation:
\begin{equation}
\Delta\phi=\arctan\left(2\pi f R C\right)
\label{e:deltaphi}
\end{equation}
where $f$ is the operational frequency, $R$ is the resistance and $\Delta\phi$ is the phase lag. By measuring the phase lag, the capacitance can be computed by:
\begin{equation}
C=\frac{1}{2 \pi f R}\tan\left(\Delta\phi\right)
\label{e:C}
\end{equation}
In practice, the phase lag is measured by the time delay between the input sine and the resulting output. The time delay $\Delta t$ can be converted to a phase lag according to the equation:
\begin{equation}
\Delta\phi=360\degree f \Delta t
\label{e:deltaphi2}
\end{equation}
The capacitance can be calculated from the measured time delay as:
\begin{equation}
C=\frac{1}{2 \pi f R}\tan\left(360\degree f \Delta t\right)
\label{e:C2}
\end{equation}
The general structure of the circuit is presented in Fig. \ref{f:meas_circuit}. The oscillator generates the input sine, the readout section converts the time delay between a reference resistor and the capacitor under test (CUT) into a square wave and the digital counter measures the ON period of the pulse generated by the readout section.  The circuit was simulated using LTSpice and then implemented on a PCB manufactured by Eurocircuits.
The oscillator to generate the input sine wave was implemented by a buffered quadrature oscillator, where each section provides a shift of -60° so that the three sections combined achieve the 
-180° necessary for the oscillation. A gain of 8.33 was necessary to sustain the oscillations. The targeted frequency was 20Hz, however, due to the mismatch between the components the circuit oscillates at 44Hz with an amplitude of 500mVpp. At the output of the oscillator a digital potentiometer, directly set by the microcontroller, was added to control the amplitude of the sine (Fig. \ref{f:meas_circuit}). To avoid the use of a dual supply, which would have required extra components on the board powered via the USB single supply, the circuit was biased at 2.5V through a Zener diode, which was used as voltage reference for the analog part of the circuit. 
The generated sine wave is sent to the RC section of the circuit, where is converted to a square wave by comparator U2. The delayed voltage across the capacitor is also converted to a square wave through comparator U3. The two resulting square waves are xored to generate a pulse that corresponds to the time delay between the two signals (Fig. \ref{f:meas_circuit}). U2 and U3 are fast comparators, whose propagation delay (4.5ns) will only add an error of 0.02 fF.
The pulse coming from the read-out section is used to gate a 12 MHz clock fed to a 64 bits counter, which will convert the pulse into a digital count. The count is then transferred to the microcontroller via a parallel to serial converter. The time delay is then calculated via the counter count as:
\begin{equation}
\Delta t=\frac{n_{count}}{12 MHz}
\label{e:deltat}
\end{equation}
This approach provides a theoretical resolution of $3 \times 10^{-16}$ F and a dynamic range of 125dB. To validate the circuit, we used different calibration ceramic capacitors (100, 300, 500, 1000 pF, 10\%, 50V) that we previously calibrated with an impedance analyzer (MFIA Zurich Instruments, accuracy 0.05\%) using the same frequency and output amplitude. The parasitic capacitance was measured by measuring the capacitance without any load ($\sim$150 pF) and then subtracted to the measured values. The results are summarized in the table below:

\begin{center}
\begin{tabular}{ |p{2.5cm}|p{2.5cm}|p{2.5cm}|}
\hline
 Nominal Value (pF) & MFIA (pF) & Capacitance Meter (pF) \\
 \hline
 100 & 105 & 103 \\  
 \hline 
 300 & 308 & 309 \\  
 \hline 
 500 & 481 & 493 \\  
 \hline
 1000 & 977 & 974 \\
 \hline    
\end{tabular}
\end{center}

The calibration shows that the capacitances measured by the circuit are within the tolerance of the nominal values of the measured capacitors and good accordance with the capacitance measured with the impedance analyzer.

\vspace{10px}
\section*{Acknowledgment}
We thank Luca Bonanomi for fruitful discussions during the initial stages of the project.

\bibliographystyle{IEEEtran}
\bibliography{ref}
\vspace{7cm}

\end{document}